\documentclass[a4paper,10pt,reqno,twocolumn]{amsart}
\usepackage[left=2cm,right=2cm,top=2.1cm,bottom=2cm]{geometry}

\usepackage[colorlinks=true, citecolor=black, linkcolor=black]{hyperref}
\usepackage{cite}

\usepackage{amsfonts}
\usepackage{amsmath}
\usepackage{amssymb}
\usepackage{bm}

\theoremstyle{remark}

\def\gt{\tilde{g}}
\def\xib{\bar{\xi}}
\def\Ecal{{\mathcal E}}
\def\Ecalb{\bar{\mathcal E}}

\def\ba{\begin{array}}
\def\ea{\end{array}}
\def\la{\label}
\def\p{\partial}

\def\f{\frac}

\newcommand{\nn}{\nonumber}

\newcommand{\be}{\begin{equation}}
\newcommand{\ee}{\end{equation}}
\newcommand{\ben}{\begin{displaymath}}
\newcommand{\een}{\end{displaymath}}
\newcommand{\bea}{\begin{eqnarray}}
\newcommand{\eea}{\end{eqnarray}}

\hypersetup{
           breaklinks=true,   
           colorlinks=true,   
           pdfusetitle=true,  
        }

\begin{document}

\twocolumn[
\begin{@twocolumnfalse}

\begin{titlepage}
\vfill

\vfill
\begin{center}
	{\LARGE \bf Integrability and Einstein's Equations}\bigskip
	
	{ \bf 
D. Korotkin\,$^{a}{\!}$
		\footnote{\tt dmitry.korotkin@concordia.ca},
H. Samtleben\,$^{b,c,}{\!}$
		\footnote{\tt henning.samtleben@ens-lyon.fr}
		 \vskip .2cm}
	
	{\it ${}^a$ Department of Mathematics and Statistics, Concordia University,
1455 de Maisonneuve West, Montreal, H3G 1M8  Quebec,  Canada}\\ \vskip .1cm
	{\it ${}^b$ ENSL, CNRS, Laboratoire de physique, F-69342 Lyon, France}\\ \vskip .1cm
	{\it  $^{c}$ Institut Universitaire de France (IUF)}\\ \ \\
	
\end{center}
\vfill

\begin{center}
	\textbf{Abstract}
	
\end{center}
\begin{quote}
Integrable structures arise in general relativity when the spacetime possesses a pair of commuting Killing vectors
admitting 2-spaces orthogonal to the group orbits.
The physical interpretation of such spacetimes depends on the norm of the Killing vectors.
They include stationary axisymmetric spacetimes, Einstein-Rosen waves with two polarizations, Gowdy models,
and colliding plane gravitational waves.
We review the general formalism of linear systems with variable spectral parameter,
solution generating techniques, and various classes of exact solutions. In the case of the Einstein-Rosen waves,
we also discuss the Poisson algebra of charges and its quantization.

This is an invited contribution to the 2nd edition of the Encyclopedia of Mathematical Physics.
\end{quote}
\vfill
\setcounter{footnote}{0}

\end{titlepage}

\bigskip
\bigskip

\end{@twocolumnfalse}
]


\setcounter{page}{1}

The theory of integrable systems and the theory of gravity, being two independent areas of research, have, however, a non-trivial intersection.
The notion of integrability itself has many facets. Its meaning varies from complete integrability in the Liouville sense to ``exact solvability" in the sense
of the existence of large classes of exact solutions which can be constructed due to the existence of the so-called Lax pair associated to a given non-linear equation.
The Liouville integrability and the exact solvability are equivalent in some cases, like the Korteveg de Vries (KdV) equation and its numerous cousins (see the classical textbooks \cite{Novikov:1984id,Babelon:2003qtg}). In Einstein gravity with sufficient number of symmetries the integrability 
is understood in the sense of ``exact solvability", or the existence of an infinite-dimensional symmetry group (the Geroch group \cite{Geroch:1972yt}).
While the full Einstein equations without symmetries are not integrable in any sense, the integrability in the above sense arises if the manifold admits two commuting Killing vectors which in turn admit 2-spaces orthogonal to the group orbits.
If one of those Killing vectors is timelike, and another one is spacelike, such spacetimes are stationary and axially symmetric.
If both Killing vectors are spacelike, there are several possibilities: the axially symmetric gravitational waves (Einstein-Rosen waves),
colliding plane gravitational waves, and the Gowdy models. 
The discussion of formal aspects of integrability is parallel in all of these cases (they differ by an appropriate Wick rotation). We shall mainly discuss the formalism in application to stationary axially symmetric spacetimes.
 
In  Weyl canonical coordinates $(t,\varphi,z,\rho)$ the metric of a stationary axially symmetric spacetime can be written as follows:
\be
ds^2 = e^{\Gamma} (d\rho^2+d z^2)+\rho\,g_{ab}(\rho,z) \,dx^a dx^b
\;,
\la{m1}\ee
where $a,b=0,1$, $x_0=t$, $x_1=\varphi$. The timelike Killing vector is then $\partial_t$ while the spacelike one is $\partial_\varphi$. The symmetric matrix $g$ satisfies ${\rm det} \,g=-1$.
Parametrizing this matrix as
\be
g=-\frac1{\rho}
\begin{pmatrix} f & fA\\[.2ex]
fA & fA^2-f^{-1}\rho^2 \end{pmatrix}
\;,
\la{g0}
\ee
the metric (\ref{m1}) takes the Lewis-Papapetrou form
(see \cite{Stephani:2003tm}, section 19.3):
\be
d s^2 = f^{-1} \left[e^{2k} (d \rho^2 +d z^2)+\rho^2  d\varphi^2\right]-f(d t+A d\varphi)^2\;,
\la{m2}
\ee
with
\be
\Gamma= 2k-\ln f
\;.
\ee
The Einstein equations imply the non-linear PDE for the matrix $g$:
\be
(\rho g_\rho g^{-1})_\rho+ (\rho g_z g^{-1})_z=0\;,
\la{eq1}
\ee
and for each $g$ satisfying (\ref{eq1}) the function $\Gamma$ can be computed in curvatures from the following compatible system \cite{BelZak2}:
\bea
\Gamma_{\rho}&=&-\rho^{-1} +\frac{\rho}{4}\, {\rm tr}  (J_\rho^2-J_z^2)\;,
\nonumber\\
\Gamma_z&=&\frac{\rho}{2} \, {\rm tr} (J_\rho J_z)
\;,
\la{eqG}
\eea
with
\be
J_\rho = \partial_\rho g \, g^{-1}\;,\quad
J_z= \partial_z g \, g^{-1}
\;.
\ee
The initial conditions for equations (\ref{eqG}) are typically chosen to provide the regularity of the metric 
(\ref{m1}) at infinity.

To give the dual form of equation (\ref{eq1}) we introduce the matrix
\be
\gt=\frac{1}{\Ecal+\Ecalb}\begin{pmatrix}  2 & -i(\Ecal-\Ecalb) \\ -i(\Ecal-\Ecalb) & 2\Ecal\Ecalb \end{pmatrix}
\;,
\la{gt}
\ee
where the complex-valued function $\Ecal(\rho,z)$ (the Ernst potential) is related to the
coefficients $f$ and $A$ of (\ref{g0}) via the equations
\be
f={\rm Re} \Ecal\;, \hskip0.7cm
\frac{\p A}{\p\xi}=2\rho \frac{(\Ecal-\bar{\Ecal})_{\xi}}{(\Ecal+\bar{\Ecal})^2}
\;,
\la{fA}\ee
where $\xi=z+i\rho$.
Then the equation (\ref{eq1}) is equivalent to the matrix equation for $\gt$
\be
(\rho \gt_\rho \gt^{-1})_\rho+ (\rho \gt_z \gt^{-1})_z=0
\;,
\la{eq3}
\ee
which formally looks identical to (\ref{eq1}).
In turn, equation (\ref{eq3}) is equivalent
to following complex scalar equation (the Ernst equation \cite{Ernst:1967wx}) for the Ernst potential~$\Ecal$:
\be
(\Ecal+\bar{\Ecal})(\Ecal_{zz}+\frac{1}{\rho}\Ecal_{\rho}+\Ecal_{\rho\rho})=2 (\Ecal_z^2+\Ecal_\rho^2)
\;.
\la{Ernst}
\ee
The function $k$ from (\ref{m2}) can be computed in terms of the Ernst potential $\Ecal$ 
by integrating the equation
\be
\frac{\p k}{\p\xi}=2i\rho \frac{\Ecal_\xi \bar{\Ecal}_{\xi}}{(\Ecal+\bar{\Ecal})^2}\;,
\la{coeff}
\ee
equivalent to (\ref{eqG}).

\section{Integrability in dimensionally reduced gravity: the $U-V$ pair with variable spectral parameter}

The equivalent equations (\ref{eq1}), (\ref{eq3}), and (\ref{Ernst}) are integrable in the sense of existence 
of the so-called $U-V$ pair, or {\it zero curvature representation}  (the generalization of the so-called Lax representation of the KdV equation \cite{Novikov:1984id})
which boils down to ``exact solvability".
Unlike for integrable systems of KdV type, here this does not imply Liouville integrability 
due to non-autonomous nature of (\ref{eq1}) and (\ref{Ernst}): the variable $\rho$ enters these equations explicitly.

Different but equivalent $U-V$ pairs for  equations (\ref{eq1}) and (\ref{Ernst}) were found in 1978 in \cite{Maison:1978es} and 
\cite{Belinsky:1971nt}, and in still another form slightly later in
\cite{Neugebauer:1980xx}. 
Before formulating these results we introduce the complex variables $\lambda$  and $\gamma$ (called the 
``constant spectral parameter" and the ``variable spectral parameter", respectively) as
\be
\gamma(\lambda,\xi,\xib)=\frac{2}{\xi-\xib}\left(\lambda-\frac{\xi+\xib}{2}+\sqrt{(\lambda-\xi)(\lambda-\xib)}\right)
\;,
\la{var}\ee
which is nothing but the uniformization map of the genus zero Riemann surface of the function
$$w=\sqrt{(\lambda-\xi)(\lambda-\xib)}\;.$$

Consider now the following linear system for the $2\times 2$ valued function $\Psi(\xi,\xib,\lambda)$:
\be
\f{\p\Psi}{\p\xi}=\f{g_\xi g^{-1}}{1+\gamma}\Psi\;,\hskip0.7cm
\f{\p\Psi}{\p\xib}=\f{g_{\xib} g^{-1}}{1-\gamma}\Psi
\;.
\la{UV1}
\ee
The non-linear equation (\ref{eq1}) then is the compatibility condition of the linear system (\ref{UV1}) 
for all values of $\lambda$.
In other words, the equation (\ref{eq1}) is the condition that the connection
$U d\xi+ V d\xib$, where 
\be
U=\f{g_\xi g^{-1}}{1+\gamma}\;,\hskip0.7cmV=\f{g_{\xib} g^{-1}}{1-\gamma}\;,
\ee 
has zero curvature, i.e.
\be
U_{\xib}-V_\xi+[U,V]=0
\;.
\ee

The original Belinskii-Zakharov $U-V$ representation is written assuming that the variables $(\xi,\xib,\gamma)$ are
independent. In these variables the derivatives in the left-hand side of equations (\ref{UV1}) become linear combinations of derivatives with respect to $(\xi,\gamma)$ and $(\xib,\gamma)$, respectively.
In the formalism of \cite{Maison:1978es} and \cite{Neugebauer:1980xx} the variables $(\xi,\xib,\lambda)$ are considered as independent, and their $U-V$ pairs are essentially equivalent to (\ref{UV1}).
In particular, the $U-V$ pair of \cite{Neugebauer:1980xx} looks as follows:
\begin{align}
\frac{\p \Phi}{\p\xi}=\frac{1}{\Ecal+\Ecalb}\left[
\begin{pmatrix}\Ecalb_\xi & 0 \\ 0 & \Ecal_\xi\end{pmatrix}
+\sqrt{\frac{\lambda-\xib}{\lambda-\xi}}
\begin{pmatrix} 0 & \Ecalb_\xi  \\ \Ecal_\xi & 0\end{pmatrix} \right]\Phi
\;,
\nonumber\\[1ex]
\frac{\p \Phi}{\p\xib}=\frac{1}{\Ecal+\Ecalb}\left[
\begin{pmatrix}\Ecalb_{\xib} & 0 \\ 0 & \Ecal_{\xib}\end{pmatrix}
+\sqrt{\frac{\lambda-\xi}{\lambda-\xib}}
\begin{pmatrix}0 & \Ecalb_{\xib}  \\ \Ecal_{\xib} & 0\end{pmatrix} \right]\Phi
\;,
\la{NKUV}
\end{align}
where $\Phi$ is a $2\times 2$ matrix function.

The $U-V$ pairs (\ref{UV1}) or (\ref{NKUV}) are the starting points for casting the non-linear differential equations
into a matrix Riemann-Hilbert problem, which is a problem of complex analysis,
and further application of various solution generating techniques. 
There are several different formulations of these Riemann-Hilbert problems. The convenient choice of such formulation
depends on the class of solutions in question and on the signs of norm of the Killing vectors.

\section{Multisoliton solutions, Geroch group and Kerr black holes}

 The multisoliton solutions of equation (\ref{eq1}) can be naturally cast into the framework of the 
 infinite-dimen\-sional Geroch group \cite{Geroch:1972yt}.
 From the point of view of integrable systems this group can be described as follows (this description was first derived in  \cite{Belinsky:1971nt},
 but we shall present it using the equivalent  linear system (\ref{NKUV})).
 Let $\Phi_0$ be a given ``seed solution" of (\ref{NKUV}) corresponding to the Ernst potential $\Ecal_0$ and satisfying the
 symmetry relation $\Phi_0(\lambda^*)=\sigma_3 \Phi_0(\lambda)\sigma_3$ where the involution $*$ changes the sign of the square root
 in (\ref{NKUV}).
 Define the new function $\Phi$ as follows:
 \be
 \Phi=T(\gamma,\xi,\xib) \,\Phi_0
 \;,
 \la{Phisol}
 \ee
 where $T=\sum_{j=-n}^n T_j(\xi,\xib)\gamma^j$ for some $n$ (the number $2n$ corresponds to the number of solitons added to the ``seed" solution). Due to the structure of the matrix of coefficients of (\ref{NKUV}) one assumes that the matrix $T$ satisfies the symmetry condition
 $T(\gamma^{-1})= \sigma_3 T(\gamma)\sigma_3$. In addition, one chooses real constants 
 $\{\lambda_j\}_{j=1}^n$ and  constants 
 $\{\alpha_j\}_{j=1}^n$ (such that $|\alpha_j|=1$) and imposes the condition that 
 ${\rm det} \,T(\lambda_j)=0$ with the null eigenvector is defined by
\be
 T(\lambda_j)\Psi_0(\lambda_j)\begin{pmatrix} 1 \\  \alpha_j \end{pmatrix}=0
 \;.
\la{solcond}
\ee
 The linear system (\ref{solcond}) for the Laurent coefficients $T_j$ 
 of the matrix $T$ 
 may be solved by Kramer's rule to give the
 determinant representation for the $2n$-soliton solution $\Ecal$ 
 on the background of the initial
 seed solution $\Ecal_0$ \cite{Belinsky:1971nt}, \cite{Neugebauer:1980xx}. 
 In the theory of integrable systems, adding multisolitons to an arbitrary seed
 solution goes under various names such as ``dressing" or ``B\"acklund" transformations.
 The constants $\lambda_j$ can also form complex conjugated pairs with appropriate modification of the reality conditions for 
 $\alpha_j$'s. 
 
 For $n=1$, applying the dressing procedure to Min\-kowski spacetime, one obtains the family of Kerr-NUT solutions, including the Kerr black hole solution itself.  
 For $n=2$ this scheme gives a family of solutions describing a superposition of two Kerr-NUT solutions \cite{NK}. 
 As was shown in \cite{Veselov:1983kp}, none of these configurations can be of physical significance due to the
 existence of conical defects and closed timelike curves on the part of the symmetry axis connecting the black holes
 (however, in the context of gravitational  waves large classes of multi-soliton solutions do not possess obvious non-physical features).

 The symmetry group generated by the dressing transformations is equivalent to the so-called {Geroch group}
 \cite{Geroch:1972yt} whose infinitesimal form was actually discovered in 1972, long before the theory of integrable systems was applied to these equations. As
 shown in \cite{Breitenlohner:1986um}, this group can be identified with the loop group $\widehat{{\rm SL}(2)}$, 
 and if one also takes into account its action on the conformal factor $\Gamma$ in (\ref{m1}), one obtains the 
 central extension of $\widehat{{\rm SL}(2)}$, \cite{Julia:1981wc}.

Although in the stationary axisymmetric case, all multisoliton solutions beyond the Kerr solution itself possess unphysical features as
long as the number of solitons remain finite, the infinite soliton chain can be interpreted as rotating black hole in a universe periodic in $z$-direction \cite{Peraza:2022xic}. Such solutions generalize the periodic Schwarzschild solutions (which are static, and therefore can be obtained by an elementary linear superposition of an infinite number of the regular Schwarzschild black holes) \cite{Myers:1986rx} \cite{Korotkin:1994dw} \cite{Frolov:2003kd}.

\section{Algebro-geometric solutions and rotating dust discs}

A more complicated class of solutions which can still be described explicitly is the class of algebro-geometric solutions found in \cite{Korotkin:1988yc}.
These solutions generalize the multi-soliton ones and can be
expressed in terms of hyperelliptic Riemann theta-functions. For traditional integrable systems of KdV -type the algebro-geometric solutions are periodic or quasi-periodic with respect to space-time variables
(see the textbook \cite{Babelon:2003qtg} for details and references); however, their degenerate limits are 
localized soliton solutions.

Let us consider the Ernst equation (\ref{Ernst}).   A special feature of algebro-geometric (also called ``finite-gap") solutions of (\ref{Ernst}) is that the underlying Riemann surface explicitly depends on the spacetime variables.  Here we discuss the simplest case of the elliptic (genus 1) spectral curve, referring to \cite{Korotkin:1988yc} and the textbook \cite{KleinRichter}. Namely, consider the elliptic curve
\be
\omega^2=(\lambda-\xi)(\lambda-\xib)(\lambda-\lambda_0)(\lambda-\bar{\lambda}_0)
\;,
\la{elc}
\ee
where $\lambda_0\in C$ is a constant. The curve (\ref{elc})  has four branch points: two of them are fixed ($\lambda_0$ and $\bar{\lambda}_0$) and two depend on the spacetime variables ($\xi$ and 
$\xib$). Consider the holomorphic differential $v=\frac{d\lambda}{\omega}$. The module of the curve (\ref{elc}) is given by the ratio of two full elliptic integrals:
\begin{equation}
\sigma=\left({ \int_{\bar{\lambda}_0}^{\lambda_0} \!v}\right)^{-1}{\int_{\xi}^{\lambda_0} \!v}
\;.
\end{equation}
Define the ratio of elliptic integrals
\begin{equation}
J=\frac12\, \left({ \int_{\bar{\lambda}_0}^{\lambda_0} \!v}\right)^{-1}
{\int_{\xi}^{\infty^+}\!\!\!v }\;,
\end{equation}
and pick a real constant $q\in R$. Consider also the Jacobi theta-function $\theta(x)=\theta_3(x,\sigma)$ associated to the curve (\ref{elc}). Then the elliptic solution of the Ernst equation can be written as
\be
\Ecal(\xi,\xib)=\frac{\theta(J+i q)}{\theta(J-iq)}
\;.
\la{ell}
\ee
When in the right-hand side of  (\ref{elc}) there are $2g$ instead of $2$ monomials independent of $\xi$ and $\xib$, 
a straightforward analog of (\ref{ell}) is expressed in terms of multi-dimensional Riemann theta-functions associated to a hyperelliptic algebraic curve of genus $g$ with one ``moving" branch cut
$[\xi,\xib]$ and $g$ branch cuts independent of $\xi$ and $\xib$. The ends of the fixed branch cuts 
can be either complex conjugate to each other or real.

When all fixed branch cuts degenerate to a point, the algebro-geometric solutions degenerate to multi-soliton ones. In particular the Kerr-NUT solution is a degeneration of the genus two algebro-geometric one \cite{Korotkin:1988yc}.

The algebro-geometric 
 solutions of the Einstein
equations are not periodic or quasi-periodic as in the KdV case. Instead they have similar asymptotic behaviour as the multi-soliton ones (i.e.\ multi Kerr-NUT solutions).

In \cite{Neugebauer:1995pm} it was shown that a special genus two algebro-geometric solution solves the boundary value problem corresponding to 
an infinitely thin relativistic rigidly rotating dust disk. See
\cite{Klein:1999zw}, \cite{KleinRichter} for applications to other potentially physically relevant boundary value problems
which correspond to disks consisting of two counter-rotating components of dust. The mathematical 
approach to boundary value problems related to algebro-geometric solutions was later formulated in \cite{LenFocas,Lenells:2010wt}.

\section{Relationship to isomonodromic deformations and Schlesinger system}

The existence of  algebro-geometric solutions of the Ernst equation is due to the general phenomenon
described in \cite{Korotkin:1994au}, namely, the intimate link between equations (\ref{eq1}) and (\ref{Ernst}) to the  theory of isomonodromic deformations
and the classical Schlesinger equations underlying these deformations \cite{Jimbo:1981zz}.
Specifically, these are isomonodromic deformations of systems of two linear differential equations with Fuchsian singularities of the type 
\be
\frac{d\Psi}{d \gamma} =\sum_{j=1}^N \frac{A_j}{\gamma-\gamma_j}\Psi
\;,
\ee
equipped with the initial condition $\Psi(\infty)=I$. 
Assuming that the monodromies of this linear system are independent 
of positions of singularities $\gamma_j$ implies that the function $\Psi$ satisfies 
the following differential equations with respect to $\gamma_j$:
\be
\frac{\p\Psi}{\p \gamma_j}=-\frac{A_j}{\gamma-\gamma_j}\Psi
\;.
\la{eqgj}
\ee
The compatibility of equations (\ref{eqgj}) with the original system implies the classical Schlesinger equations for the residues $A_j$ with respect to positions of poles $\gamma_k$.

The relationship of the theory of isomonodromic deformation to Einstein's equations stems from the 
following  observation \cite{Korotkin:1994au}: suppose the number of poles $\gamma_j$ is even, and they are 
split into pairs formed by $\gamma_j=\gamma(\lambda_j,\xi,\xib)$ and $\gamma_j^{-1}$. If one further assumes that the corresponding monodromies are given by $M_j$ and $\sigma_3 M_j \sigma_3$ for 
arbitrary $M_j$ such that the product of all monodromies is $I$, then the function $\Psi$ satisfies the linear system (\ref{UV1}) for some $g$. Thus, such $g$ (which can in turn be expressed via the solution of the Schlesinger system) solves the Einstein equations (\ref{eq1}).

The multi-soliton and algebro-geometric solutions are special cases of this general construction: 
for multisoliton solutions all monodromies are trivial (equal to $I$) and $\Psi$ is a rational function of $\gamma$.
For algebro-geometric solutions some monodromies are off-diagonal while the others are diagonal
(one can take the limit when the number of the latter tends to infinity, see \cite{Korotkin:1988yc}).

The problem of finding the function $\Psi$ for a given set of monodromies is called the matrix Riemann-Hilbert problem. The above observation means that each explicit solution of the Riemann-Hilbert problem can be used to construct an explicit solution of Einstein's equations.

The key ingredient of the theory of isomonodromic deformations is the Jimbo-Miwa tau-function which is the scalar function whose zero locus is related to the set of solvable Riemann-Hilbert problems.
As it was shown in \cite{Korotkin:1994au} the tau-function turns out to coincide (up to an elementary factor) with the conformal factor
$e^\Gamma$ from (\ref{m1}) under the above correspondence.

\section{Self-dual Einstein metrics}

Self-dual Einstein metrics of Euclidean curvature can also be studied by methods originating in the theory of integrable systems,
or the closely related twistor theory \cite{Mason:1991rf}. Namely, the self-dual Einstein equation, also called Plebanski's heavenly equation,
can be studied by the same methods as various dispersionless integrable system. For example, the self-dual Einstein equations 
with one Killing vector (the Boyer-Finley equation $U_{xy}=(e^U)_{tt}$ \cite{Boyer:1982mm}) is, strictly speaking, not integrable; it possesses neither
a complete family of integrals of motion nor a zero curvature representation. Still, in analogy to the dispersionless Kadomtsev-Petviashvili (KP) equation, some classes of  solutions can be constructed via the so-called generalized hodograph method used to solve systems of hydrodynamic type, see \cite{Calderbank:1999ad,Ward:1990qt,Dunajski:2000rf,Manas:2004,Ferapontov:2002jk}.

\subsection*{Spherically symmetric self-dual Einstein equations: Bianchi  models and Painlev\'e equations}

In the simplest case, when the metric possesses ${\rm SU}(2)$ invariance, the Euclidean Einstein equations
 with cosmological constant
 can be analyzed by the
usual methods of integrable systems.
The Einstein equations then reduce to the so-called Painlev\'e equations 
which are special cases of the $2\times 2$ Schlesinger systems.
Consider the following form of an ${\rm SU}(2)$ invariant Euclidean metric \cite{Tod:1994}:
\be
ds^2= F\left\{d\mu^2 +\frac{\sigma_1^2}{W_1^2}+\frac{\sigma_2^2}{W_2^2}+\frac{\sigma_2^2}{W_2^2}\right\}
\;,
\ee
where the 1-forms $\sigma_j$ satisfy $d \sigma_1=\sigma_2\wedge d\sigma_3$, etc., and the functions $W_j$ 
depend only on Euclidean time $\mu$.
Defining the connection coefficients $A_j$ via the relations
\be
\frac{d W_j}{d \mu}= -W_k W_l +W_j(A_k+A_l)
\;,
\la{conn}\ee
where $(j,k,l)$ is any permutation of $(1,2,3)$,
the Einstein equations imply the following system of equations for $A_j$ (due to Halphen):
\be
\frac{d A_j}{d \mu}= -A_k A_l +A_j(A_k+A_l)\;.
\label{Halphen}
\ee
The solution of these equations can be written in terms of Jacobi's theta-constants as follows:  
\be
A_j=2\frac{d}{d\mu} \ln \theta_{j+1}(0,i\mu)\;,
\label{solHal}
\ee
 where $\theta_2$, $\theta_3$ and $\theta_4$ are  Jacobi's theta-constants.
 The general solution can be obtained by applying to this solution a M\"obius transformation of $\mu$.
Then the  equations (\ref{conn}) for the metric coefficients turn out to be equivalent to the special explicitly solvable case of the Painlev\'e 6 equation \cite{Tod:1994}.
The resulting formulas for $W_j$ and $F$ can be also nicely represented in terms of Jacobi theta-functions \cite{Hitchin:1995hxv},\cite{Babich:1998tz}. 

The general class of solutions (\ref{solHal}) of (\ref{Halphen}) corresponds to metrics of Bianchi IX type. The system 
(\ref{Halphen}) admits also special classes of solutions (for example, when all $A_j=0$). The corresponding metrics 
belong to other Bianchi classes, see \cite{Pederson:1990,Eguchi:1978xp,Tod:1994}.

\section{Cylindrically symmetric gravitational waves: classical and quantum Yangian structures}

The line element for cylindrically symmetric gravitational waves
(the Einstein-Rosen waves) can be obtained by simultaneous Wick rotation of variables $t$ and $z$ in the line element of stationary axially symmetric spacetimes. This line element is written in the form
\be
ds^2 = e^{\Gamma(\rho,\tau)} (-d \tau^2 + d\rho^2) +\rho\,{g}_{ab} (\tau,\rho) \,
dx^a dx^b\;, \la{m}
\ee
where $a,b=2,3$, $x^2=z$, $x^3=\varphi$, with radial coordinate $\rho$
and time $\tau$. In this case, the Killing vectors are both spacelike, given by $\partial_\varphi$ and $\partial_z$. 
The symmetric $2\times 2$ matrix ${g}(\tau,\rho)$ 
satisfies the condition ${\rm det}\;{g}\!=\!1$.

The Einstein equations now reduce to
\be
(\rho g_\rho g^{-1})_\rho- (\rho g_\tau g^{-1})_\tau=0
\;,
\la{eq5}
\ee
and the following analog of equations (\ref{eqG}) for $\Gamma$:
\bea
 \Gamma_{\rho}&=&-\rho^{-1} +\frac{\rho}{4} \,{\rm tr}  (J_\rho^2+J_\tau^2)\;,
\nonumber\\
\Gamma_\tau&=&\frac{\rho}{2}  \,{\rm tr} (J_\rho J_\tau)
\;,
\la{eqG1}
\eea
where
\be
J_\rho=g_\rho g^{-1}\;,\hskip0.7cm J_\tau=g_\tau g^{-1}
\;.
\label{jj}
\ee
Equations (\ref{eq5}) can be derived from the Lagrangian
\be
\mathcal{L}^{(2)}(\rho,\tau) ~=~  \frac1{2G}\,\rho\,
{\rm tr} \Big(J_\rho^2 - J_\tau^2\Big)\;,
\la{Lag}
\ee
which arises from the full 4d Einstein Lagrangian in the process of dimensional reduction.

The associated linear system is obtained by Wick rotation from (\ref{UV1}):
\be
\f{\p\Psi}{\p x_\pm }=\f{g_{x_\pm} g^{-1}}{1\pm \gamma}\Psi\;,\hskip0.7cm
\la{UVpm}
\ee
where $x_\pm = \tau\pm \rho$, and the variable spectral parameter is given by
\be
\gamma(\lambda,x_+,x_-)=-\frac{1}{\rho}\left(\lambda-\tau +\sqrt{(\lambda-\tau)^2-\rho^2}\right)
\;,
\la{varpm}
\ee
and lives on the Riemann surface defined by the function $\sqrt{(\lambda+\tau+\rho)(\lambda+\tau-\rho)}$.
The non-linear equation (\ref{eq5}) is the compatibility condition of the linear system (\ref{UVpm}).

From the solution $\Psi$ of the linear system (\ref{UVpm}), one 
defines the transition matrices
\bea
T_\pm(\lambda,\tau)&=& \Psi(\rho=0, \gamma(\lambda),\tau) \Psi^{-1}(\rho=\infty, \gamma(\lambda),\tau) 
\;,
\nonumber\\
&& \mbox{for } \Im\lambda \gtrless 0
\;,
\la{Tpm}
\eea
defined as holomorphic functions of $\lambda$ in the upper and the 
lower half of the complex plane, respectively. In (\ref{Tpm}) the variable spectral parameter $\gamma$ 
is chosen on the branch inside the unit circle, i.e.\ $|\gamma|<1$.
Definition (\ref{Tpm})
further implies $\det T_\pm\!=\!1$ and
$T_+(\lambda)=\overline{T_-(\bar{\lambda})}$.
Assuming that the physical currents $J_\rho$, $J_\tau$, from (\ref{jj})
fall off sufficiently fast at spatial infinity $\rho\rightarrow\infty$,
the matrices $T_\pm$ are constants of motion, i.e. 
\be
\p_\tau T_\pm(\lambda,\tau) = 0\;.
\ee
Generically, the matrices $T_\pm$ do not coincide in the limit to the real $w$-axis. 
Their  product
$M=T_+T_-^\top$ (called the monodromy matrix in \cite{Breitenlohner:1986um}) on the real axis has a well-defined physical meaning,
namely it coincides with the values of the original matrix $g$ on the
symmetry axis:  
\be
M(\lambda\!\in\!\mathbb{R})\equiv T_+(\lambda) T_-^\top(\lambda) = 
g(\rho\!=\!0, \tau\!=\!\lambda)\;.
\label{M}
\ee
In particular, it is symmetric and real:
\be\la{con}
M(\lambda) = M^\top(\lambda) \qquad {\rm and } \qquad  
M(\lambda) = \overline{M(\lambda)}  ~.
\ee
Since the $T_\pm$ contain the initial values of the metric and the
Ernst potential on the symmetry axis $\rho\!=\!0$, they contain
sufficient information to recover $g$ everywhere by means of
equations of motion  (note that $\p_\rho g(\rho\!=\!0)=0$ for solutions regular on
the symmetry axis). Thus, the set of $T_\pm(\lambda)$ is a 
complete set of observables for the Ernst equation.

The symplectic structure on these objects can be derived starting from 
the Lagrangian (\ref{Lag}) and its canonical equal-$\tau$ Poisson brackets
\be
\Big\{g_{ab}(\rho), (g^{-1} \p_\tau g g^{-1})_{cd}(\rho') \Big\} 
~=~\frac{G}{\rho}\,\delta_{ad}\delta_{bc}\,\delta(\rho-\rho')\;.
\label{PB}
\ee
The restrictions of symmetry and unit determinant of $g$ can be straightforwardly implemented
upon proper parametrization of the matrix.
The Poisson structure (\ref{PB}) induces the following quadratic Poisson brackets on the matrix 
entries of~$T_\pm$  \cite{Korotkin:1997fi}:
\begin{align}
&\Big\{T^{ab}_\pm(\lambda),T^{cd}_\pm(\mu)\Big\} = \nn\\
&\quad =\frac{G}{\lambda-\mu}\:
\Big(T^{ad}_\pm(\lambda)T^{cb}_\pm(\mu)-T^{cb}_\pm(\lambda)T^{ad}_\pm(\mu)\Big)\;,
\la{pa1}\\[1ex]
&\Big\{T^{ab}_-(\lambda),T^{cd}_+(\mu)\Big\} = \nn\\
&\quad = \frac{G}{\lambda-\mu}\:
\Big(T^{ab}_-(\lambda)T^{cd}_+(\mu)-T^{cb}_-(\lambda)T^{ad}_+(\mu) \nn\\
&\qquad\qquad\qquad  -\delta^{bd}\,T^{am}_-(\lambda)T^{cm}_+(\mu)\Big)\;.\la{pa2}
\end{align}
The proper quantum analogue of the Poisson brackets
(\ref{pa1}) is known as the so-called ${\mathfrak{sl}}(2)$-Yangian 
algebra  \cite{Drinfeld:1985rx}
\begin{align} \la{y1}
&\Big[T^{ab}_\pm(\lambda),T^{cd}_\pm(\mu)\Big] = \nn \\
&\quad=
\frac{i\hbar G}{\lambda-\mu}\:
\Big(T^{cb}_\pm(\mu)T^{ad}_\pm(\lambda)-T^{cb}_\pm(\lambda)T^{ad}_\pm(\mu)\Big)\;.
\end{align}
The consistent quantization  of the Poisson brackets (\ref{pa2}) 
and the symmetry relation (\ref{con}) is uniquely given by the following set of
mixed relations \cite{Korotkin:1997ps}
\begin{align}
&\Big[T^{ab}_-(\lambda),T^{cd}_+(\mu)\Big] = \nn \\
&\quad =\frac{i\hbar G}{\lambda\!-\!\mu\!+\!i\hbar G}\:T^{cd}_+(\mu)T^{ab}_-(\lambda)\la{y2} \nn\\
&{}\quad - \frac{i\hbar G(\lambda\!-\!\mu)}{q(\lambda,\mu)}\: 
\Big(T_+^{ad}(\mu)T_-^{cb}(\lambda)+\delta^{bd}\,T_+^{cm}(\mu)T_-^{am}(\lambda)\Big)\nn\\
&\quad
+ \frac{(i\hbar G)^2 }{q(\lambda,\mu)}
\;\delta^{bd}\Big(T_+^{am}(\mu)T_-^{cm}(\lambda)-T_+^{cm}(\mu)T_-^{am}(\lambda)\Big)\;,
\end{align}
where
\be
q(\lambda,\mu)=(\lambda\!-\!\mu\!+\!i\hbar G)(\lambda\!-\!\mu\!-\!i\hbar G)\;,
\ee
and the symmetry condition
\be\la{qcon}
M(\lambda) \equiv T_+(\lambda)T_-^\top(\lambda) = T_-(\lambda)T_+^\top(\lambda) 
\;.
\ee
Apart from the proper ordering of the quadratic expressions and the
quantum corrections of order $\hbar^2$ in (\ref{y2}), the essential
content of these relations is the shift of the denominator on the
r.h.s.\ in (\ref{y2}). This provides a central extension of (\ref{pa2}), \cite{Reshetikhin:1990sq},
which is required for consistency of this quantum model.
Finally, the classical
condition of unit determinant $\det T_\pm(\lambda)=1$ requires quantum
corrections because of the nonlinear terms and is substituted by the
``quantum determinant'' \cite{Izergin:2009yc,Kulish:1981bi}
\be\la{qdet}
T_\pm^{11}(\lambda\!+\!i\hbar G)T_\pm^{22}(\lambda)-
T_\pm^{12}(\lambda\!+\!i\hbar G)T_\pm^{21}(\lambda)~=~1\;,
\ee
which is indeed compatible with the relations (\ref{y1}), (\ref{y2}).
The definition (\ref{qcon}) of $M(\lambda)$ ensures 
that the commutation relations (\ref{y1}), (\ref{y2})
yield a {closed} commutator algebra of the matrix entries of
$M(\lambda)$. Moreover, these are hermitean operators, provided that 
\be\la{herm}
T_+^{ab}(\lambda) = \Big(T_-^{ab}(\bar{\lambda})\Big)^\dagger \;,
\ee
in accordance with the classical relations.  
The problem of construction of unitary representations of the quantum algebra (\ref{y1}), (\ref{y2}), (\ref{herm})
remains essentially open. A bootstrap approach to this problem was
developed in
\cite{Niedermaier:1999bh}.

More recently it was shown in \cite{Fuchs:2017jyk,Peraza:2019agr} that the Poisson algebra 
(\ref{pa1}), (\ref{pa2}) naturally arises in the Lagrangian formulation of full Einstein gravity
when the initial value problem is formulated on null surfaces.
The significance of the corresponding quadratic quantum algebra thus goes far beyond
the quantization of the dimensionally reduced gravity models.

The Poisson algebra of  the $T_\pm$ is also closely related to the 
infinite-dimensional symmetry group of equation (\ref{eq5}) (the  Geroch group). 
Specifically, the infinitesimal action of this group
is generated by Lie-Poisson action of $T_\pm$ \cite{Korotkin:1996fx, Korotkin:1997fi}.

\subsection*{Collinear polarizations.} 
Among the simplest nontrivial metrics
in the class (\ref{m}) are the collinearly polarized gravitational waves
originally discovered by Einstein and Rosen. They correspond to a
diagonal form of the matrix $g\equiv{\rm diag}(e^\phi,\,e^{-\phi})$,
i.e.\ the number of degrees of freedom reduces to one. 
Equation (\ref{eq5}) in this case reduces to the cylindrical wave
equation
\be
-\p^2_\tau\phi+\rho^{-1}\p_\rho\phi + \p^2_\rho\phi=0 \;,
\ee
with general solution
\be
\phi(\rho,\tau)=\!\int_0^\infty\!\!\!\!d\zeta\left[A_+(\zeta)
J_0 (\zeta\rho) e^{i \zeta\tau}\!+\!  
A_-(\zeta)J_0 (\zeta\rho) e^{-i \zeta \tau}\right] ,
\ee
where $J_0$ denotes the Bessel function of the first kind. The coefficients
$A_+\!=\!\overline{A_-}$ build a complete set of observables with
canonical Poisson brackets 
\be
\Big\{A_+(\zeta),\; A_- (\zeta')\Big\} = G\,\delta (\zeta-\zeta')~.
\la{psA}\ee
Thus, quantization of this structure is straightforward 
and gives rise to a representation in terms of creation and
annihilation operators 
\be\la{fock}
A_- |0\rangle = 0 \qquad {\rm with} \quad A_+=A_-^\dagger ~. 
\ee
In particular, coherent quantum states may be constructed in the same
way as in flat space quantum field theory \cite{Ashtekar:1996bb}. Historically, the quantization of this model was first performed in \cite{Kuchar:1971xm}.

To make contact with the general two polarizations case one may introduce the variables 
\be
t_\pm (\lambda) ~\equiv~\exp\int_0^\infty\!\!\!d\zeta\;  A_\pm (\zeta) e^{\pm i\lambda \zeta}\;,
\la{Tpma}\ee
which build an equivalent complete set of observables. In the Fock
space representation (\ref{fock}), $t_-(\lambda)$ is represented as the identity,
while $t_+(\lambda)$ generates the coherent state associated to a
classical field that on the symmetry axis $\rho\!=\!0$ is peaked as a
$\delta$-function at $\tau_0\!=\!\lambda$. In terms of $t_\pm$, the
Poisson structure (\ref{psA}) becomes
\be
\Big\{t_- (\lambda), t_+ (\mu)\Big\} ~=~ -\frac{G}{\lambda - \mu}\,t_-(\lambda) t_+(\mu) ~.
\la{TTa}\ee
This quadratic form of the Poisson
brackets naturally embeds into the general case of two polarizations (\ref{pa2}). 
Linearization to (\ref{psA}) is a special feature of the
truncated model but not possible in the general case. 

For a comprehensive review of quantization of midi-superspace models we refer to 
\cite{BarberoG:2010oga}.

\section{Collision of plane gravitational waves}

Another physical context where a hyperbolic version of the Ernst equation arises is the collision of two plane 
gravitational waves. Special solutions of this kind were found in \cite{Khan:1971vh}
(the Khan-Penrose solution can be obtained from Schwarzschild black hole by a
Wick rotation) and \cite{Nutku:1977wp}. The latter solution is obtained by Wick rotation from the Kerr-NUT solution. The problem was studied more systematically
using the integrable systems techniques in \cite{Hauser:1990}, among others, see 
\cite{Alekseev:2001tx} and  the book
\cite{Griffiths:1991zp} for details. More recently, a rigorous mathematical analysis 
of this problem was carried out in \cite{Lenells:2018fej}.

 The hyperbolic  version  of Ernst equation relevant in this context reads:
 \be
 (\Ecal+\bar{\Ecal})\left(\Ecal_{xy}-\frac{\Ecal_x+\Ecal_y}{2(1-x-y)}\right)=2\Ecal_x\Ecal_y
 \;.
 \label{hypE}
 \ee
 Mathematically, the problem of describing the gravitational field in the region of interaction of two plane gravitational waves with given profiles is the Goursat problem of finding the
 solution of (\ref{hypE}) in the triangle $x>0,\;y>0$, $x+y<1$, satisfying the boundary conditions
 $\Ecal(x,0)=\Ecal_1(x)$ for $x\in [0,1)$ and  $\Ecal(0,y)=\Ecal_2(y)$ for $y\in [0,1)$.

 While in the case of Einstein-Rosen waves the jump matrices of the Riemann-Hilbert problem can be derived directly from the the boundary values of the metric, in the case of plane waves these jump matrices need to be found from an integral equation involving 
 the boundary values $\Ecal_1$ and $\Ecal_2$. Once these jump matrices are 
 found and the Riemann-Hilbert problem is formulated, there remains the problem of
 existence and uniqueness of its solution for given classes of $\Ecal_1$ and $\Ecal_2$ 
 which was discussed in detail in 
 \cite{Lenells:2018fej}.
 
 \section{Integrability in models of Gowdy type}
 
 Another possible Wick rotation of the variables in equation (\ref{eq1}) (or equivalently the Ernst equation (\ref{Ernst}) is to make $\rho$ a time-like coordinate, while $t$ becomes the space-like one.
 If in addition one assumes that the metric is periodic in the $z$-direction, and considering that it is independent of $t$ 
 and also periodic in the $\varphi$  coordinate, one can see that  the space slice  has the  topology of $T^2\times \mathbb{R}$,
 which upon further compactification of the $\mathbb{R}$ factor becomes $T^3$. Models of this type with various topologies of the
 space slice are known as cosmological models of Gowdy type \cite{Gowdy:2014}. Therefore,  one can apply the 
 methods of integrable systems to  Gowdy models; however, we are not aware of published works in this direction.

\section{Matter coupling and higher coset models}

Integrability remains  when the Einstein equations are coupled to special types of matter possessing the same space-time symmetries. An example of such system is  the system of Einstein-Maxwell equations. In the stationary axisymmetric case a physically important configuration of
this type is the Kerr-Newman solution describing the charged black hole.

As far as solution-generating techniques are concerned, formally the equations of motion for such matter-coupled gravity 
in the axisymmetric case take the form (\ref{eq1}) or (\ref{eq3}), however with the matrices $g$ and $\tilde{g}$ living
on larger spaces.
In the vacuum case, the matrix $g$ in (\ref{eq1}) can be represented as $g=V\sigma_3 V^\top$,
where $V$ is a representative of 
the coset space ${\rm SL}(2,\mathbb{R})/{\rm SO}(1,1)$. Similarly, the matrix $\tilde{g}$ in (\ref{eq3}) can be represented as
$\tilde{g}=\tilde{V} \tilde{V} ^\top$ where $\tilde{V} $ is a the representative of 
the coset space ${\rm SL}(2,\mathbb{R})/{\rm SO}(2)$.
The  Einstein-Maxwell equations in the stationary axisymmetric case can be cast into the form of (\ref{eq1}) 
with $g$ corresponding to  the coset space 
${\rm SU}(2,1)/\left({\rm SU}(1,1)\times {\rm U}(1)\right)$. 

For a complete list of relevant coset models we refer to  \cite{Breitenlohner:1987dg,Cremmer:1999du}.
For different bosonic matter couplings they exhaust the classical and the exceptional groups.
The associated linear systems of the form (\ref{UV1}) are then applicable for general coset spaces. For
supersymmetric models they may be extended to also include the fermionic matter sectors \cite{Nicolai:1987kz}.
The solution generating techniques can be applied to all these cases with increasing technical complexity.

For other physical contexts (waves of Einstein-Rosen type, interaction of plane waves)
the application of solution generating techniques is parallel to the stationary axisymmetric case 
since these cases differ from the stationary axisymmetric one only by an appropriate Wick rotation.
Historically, the solution generating techniques were first applied to Einstein-Maxwell case 
in \cite{Kinnersley:1977pg}; the multisoliton solutions were given in \cite{Neugebauer:1983dp}
and \cite{Alekseev:1980ew}.
We also refer to the review \cite{Alekseev:2010mx} and Section 34.8 of \cite{Stephani:2003tm} 
for the detailed history of integrability 
of the Einstein-Maxwell equations and further technical details.


\end{document}